\documentclass[%
 preprint, 
 amsmath,amssymb,
 aps, physrev,
]{revtex4-2}

\usepackage{graphicx}% Include figure files
\usepackage{dcolumn}% Align table columns on decimal point
\usepackage{bm}% bold math

\begin{document}
% \preprint{APS/123-QED}
\title{The Three-Dimensional Velocity Field of Kinesin-Driven Microtubules in Torroidal Channels}

\author{Yi Fan}
\affiliation{Center for Fluid Mechanics, School of Engineering, Brown University, Providence RI 02912 U.S.A.}

\author{Kenneth S. Breuer}
\email{kenneth\_breuer@brown.edu}
\affiliation{Center for Fluid Mechanics, School of Engineering, Brown University, Providence RI 02912 U.S.A.}

\date{\today}

\begin{abstract}
We study two regimes of flow in multiple three-dimensional toroidal channels by tracking the fluorescent spherical particles in the kinesin-driven microtubule systems: ``chaotic'' flow and ``coherent'' flow. In the smallest aspect ratio torus, where the channel height $h$ is a quarter of the width $w$, the active system shows zero mean velocity, small-scale isotropy in fluctuation and no persistent flow structure. In other tori with higher aspect ratios $h/w$ close to 1, we find faster coherent flows along the azimuthal direction and increasing fluctuation strengths with growing confinement geometries. Regardless of flow regimes, the flow profiles at $r-z$ cross-section and $r-\theta$ plane are symmetric. The ``coherent'' profiles show two criteria: ``Poiseuille-like'' profiles, which have the peak velocities near the centers of channels; a ``peak-separated'' profile, which has four peak velocities near a certain distance to four confining surfaces. These flow profiles, after scaled by the local isotropic fluctuation strength, reveal universal three-dimensional flow structures among the ``Poiseuille-like'' criterion and the same level of scaled peak velocity at the ``peak-separated'' one. These results illustrate scalable flow structures in this kinesin-driven microtubule active system. 
\end{abstract}

\maketitle

\section{Introduction}
Active matter systems are comprised of agents that convert free energy at the smallest scales into systematic motion that could be observed at much larger scales. It covers various types of biological materials, driven by internal active stresses, and can be extended to a broader context of self-propelled particles powered by external stresses \cite{marchetti2013hydrodynamics, needleman2017active}. Among many topics that have been investigated, confinement effects provide insights into the fundamental principles of collective behavior and self-organization across a wide range of length scales. Collective motion (such as cytoplasmic streaming or rotational flows) and self-assembled structures (such as asters, spirals, vortices, etc.), introduced by boundary surfaces, have been investigated in the cytoplasm \cite{theurkauf1994premature, serbus2005dynein, schroeder1985spiral}, in bacterial suspensions \cite{di2010bacterial, sokolov2010swimming, lushi2014fluid}, in cells \cite{riedel2005self, doxzen2013guidance}, in synthetic cytoskeletal extracts \cite{ndlec1997self, schaller2010polar, suzuki2017spatial, opathalage2019self} and in self-propelled particles \cite{narayan2007long, chate2008collective}. Our attention focuses on a specific type of cytoskeletal extract, consisting of microtubules, kinesin clusters and depletion polymers. Microtubules form bundles under the depletion force from polymer micelles. The cooperation of motor clusters converts chemical energy to a sliding force between anti-aligned neighbor microtubules, and this sliding motor force gives rise to extensile flows \cite{vale1985identification, sanchez2012spontaneous}. The artificially assembled system is termed an ``active fluid'' to indicate the predominant liquid-like property. Recent experiments have elucidated two dynamical states of kinesin-driven active fluids: chaotic bulk dynamics and flows exhibiting long-range, large-scale coherence \cite{wu2017transition}. Similarly, in a recent theoretical study on the same active systems \cite{varghese2020confinement}, the 3D flow was killed and restored by varying the aspect ratio of the confining geometry. Transitions between these two dynamical states emerge in three-dimensional confinement, from micrometer to millimeter scale. The experiments also found that the large-scale coherence correlates with the nematic order extended from the confining boundaries, similar to the ``wetting'' phenomenon in liquid crystals near boundary surfaces \cite{jerome1991surface}. However, there are many issues that remain unclear. For example, the three-dimensional structure in active flows and the influence of confining boundary surfaces are unknown, and we do not yet understand why chaotic-to-coherent transition appears for specific confinement aspect ratios across a wide range of confining geometries. Lastly, the internal structure - the velocity field - of the three-dimensional coherent flow has not been described, and it is this question that forms the focus of the current work.

To study the three-dimensional flow structure, we use an experimental set-up that allows long-duration volumetric imaging of passive tracer particles using a scanning microscope objective and the out-of-focus particle tracking method \cite{fan2021effects, afik2015robust}. With this tool, we measure three components of velocity in a series of microfluidic channels with torroidal geometry. We vary the channel size and the aspect ratio, as well as the surface chemistry. In this manuscript, we report on the structure and scaling of three-dimensional velocity fields in both the chaotic and coherent flow regimes. 

\begin{figure}
    \centering
    \includegraphics[width=\columnwidth]{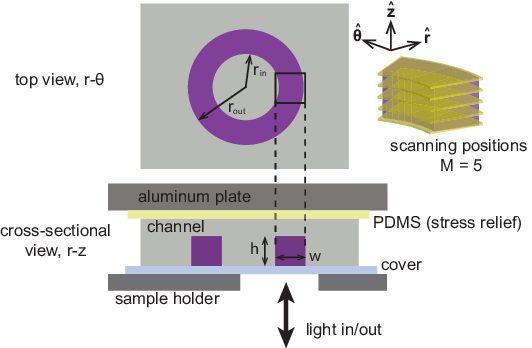}
    \caption{Schematic of toroidal test channel and test assembly. Using five ($M$ = 5) $z-$scanning positions as an example. }
    \label{fig:experiment}
\end{figure}

\begin{table*}
    \caption{Geometry and details of test channels and experimental measurements.}
    \vspace{3mm}
    \label{tab:toroids}
    \scalebox{1.0}{
    \begin{tabular}{ c | c | c | c | c | c | c | c}
        \toprule
        Material & Abbreviation & $w$ [$\mu$m] & $h$ [$\mu$m] & Aspect Ratio, $h/w$ & Coating & Cover &  Z-slices, $M$ \\ 
        \hline
        {Silicon} & SI I & 325$\pm$1.2 & 82$\pm$0.9 & 0.25 & polyacrylamide & glass & 3\\
        & SI II & 111$\pm$0.6 & 107$\pm$0.7 & 0.96 & polyacrylamide & glass & 5\\ & SI III & 218$\pm$0.7 & 206$\pm$0.3 & 0.94 & polyacrylamide & glass & 5\\
        & SI IV & 323$\pm$1.2 & 306$\pm$1.3 & 0.95 & polyacrylamide & glass &  5 \\
        \hline
        COC & COC & 298$\pm$2.2 & 292$\pm$1.2 & 0.98 & pluronic & COC & 5\\
         \hline
     \end{tabular}}
\end{table*}

\section{Methods and Materials}
\subsection{Active fluid system}
The active fluid system here has been extensively used in previous studies \cite{sanchez2012spontaneous, henkin2014tunable, wu2017transition} and consists of microtubules, kinesin-streptavidin motor clusters, depletion polymers, and the adenosine triphosphate (ATP) regeneration system. In this experiment, the microtubules were flourescently labelled and flourescent particles were added as passive tracers. Both microtubules and passive tracers were imaged and recorded. Here we only consider the motion of the passive tracers.

The active fluids were confined in a series of toroidal channels with the same inner radius, $r_{in} = 1$ mm, but varying in width $w$ and height $h$ (Fig. \ref{fig:experiment}, Table~\ref{tab:toroids}). The channels were manufactured either by etching silicon wafers or by CNC-machining cyclic olefin copolymer (COC) plates. For the experiments in the fabricated silicon channels (Table~\ref{tab:toroids}, SI I - IV), the silicon surface and the sealing cover glass were coated with polyacrylamide. The COC channel, which has similar dimensions to the silicon channel SI IV, and the sealing COC film were coated with pluronic F-127. Further details are provided in Supplementary Materials.

The active fluid was loaded into the channel and sealed by bolting together a microfluidic device (Fig. \ref{fig:experiment}) which included, from bottom to top, a sample holder to fit the motorized microscope stage (Prior Scientific, H101A), a treated cover (microscope cover glass or COC film), a treated channel (silicon channel or COC channel), a polydimethylsiloxane (PDMS) layer, $\sim$1 mm thick, to provide stress relief and an aluminum plate. The assembled package could prevent the sample from drying out for more than a week.

\subsection{Imaging and three-dimensional scanning}
The test cell was mounted on an epi-fluorescent microscope (Nikon TE 200 eclipse). The particles and microtubules were illuminated with a high power LED source (Thorlabs, DC4100) at $405$ nm and $565$ nm. The excitation light, refined using a two-bandwidth excitation filter (Chroma Technology, 59003x, $395/30$ nm and $575/30$ nm), was directed through an air immersion objective (Nikon CFI Plan Apo VC $20\times$, WD = 1 mm, NA = 0.75) mounted on a piezo objective drive (Physik Instrumente, Model P-725), through the floor of the test cell and into the test chamber. The fluorescently-labelled microtubules and fluorescent tracer particles in the sample emit light at red and blue wavelengths respectively. The emission light was passed through a two-bandwidth dichroic filter (Chroma Technology, 59003bs, transmission in $460/30$ nm and $630/60$ nm) and was directed to an optical splitter which separates the two colors using a dichroic mirror (Chroma Technology, $565$dcxr). Each light stream was passed through a narrow band filter (Chroma Technology, ET$450/50$m and D$620/60$m) and projected onto two halves of an sCMOS camera (PCO edge 5.5). Although the full sensor size of the camera is $2560$ pixels $\times$ $2160$ pixels, limitations with the windows of the optical splitter result in a final image of $2500$ pixels $\times$ $1300$ pixels. The measurement field of view in the $x-y$ plane is approximately 350 $\mu$m $\times$ 350 $\mu$m which represents around $20^{\circ}$ portion of the torroidal channel (Fig. \ref{fig:experiment}).

To achieve volumetric imaging of the flow, the focal plane was scanned over a $M$ distinct $z$-heights, or ``slices'' (Table~\ref{tab:toroids}) equally-spaced approximately $30$ $\mu$m to $70$ $\mu$m apart based on the channel height. From this $z$-stack, a quasi-steady volumetric snapshot of the particles could be assembled. The objective was moved to its location during $50$ ms, and held in place for $50$ ms (to ensure the position had stabilized) before each image was acquired with a 25 ms exposure time. In this manner a complete $z$-scan was comfortably achieved at 1 Hz for $M=3, 5$ slices. Given that the average particle velocity is approximately 2 $\mu$m/s, the slew error between different slices is minimal and can be ignored.  From the acquired images, the $x-y-z$ position of each particle, was determined from the $x-y$ position and size of the diffractive Airy ring \cite{fan2021effects}. The three components of velocity $u_r, u_\theta$ and $u_z$ are calculated from the time-resolved particle trajectories (for more details, see Supplementary Materials). 

\section{Results}
In agreement with previous experimental and theoretical results \cite{wu2017transition, varghese2020confinement}, two flow regimes were observed in the torroidal geometry, depending on the channel aspect ratio, $h/w$. In all cases, the radial and vertical velocities, $u_r$ and $u_z$, have unsteady fluctuations with zero mean. In shallow channels - low aspect ratio, $h/w$ - the azimuthal velocity, $u_\theta$, also exhibits unsteady motions with zero mean (Fig~\ref{fig:lifetime}A). This is referred as ``chaotic'' or ``turbulent'' flow regime. However, in channels with an aspect ratio, $h/w$, close to 1, a steady circulation spontaneously emerges (Fig~\ref{fig:lifetime}B,C) with a mean azimuthal velocity that peaks at 2 $\mu$m/s and decaying slowly over time. This is referred as ``coherent'' flow regime. The lifetime of the active flow in SI IV channel was more than 9 hours, while the COC channel lasted approximately 3 hours (Fig~\ref{fig:lifetime}B,C).

\subsection{Velocity of active flows}
We focus our analysis on the period $T=[0-2000]$ seconds (highlighted zone in Fig. \ref{fig:lifetime}) during which time there is minimal decay in the system activity. We decompose the velocities, $u_i(t)$ where  $i = r, \theta, z$, into two components: the time-averaged velocity, $\overline{u}_i(r, \theta, z)$,  and the fluctuation $u_i'(t) = u_i(t) - \overline{u}_i$. The strength of the fluctuation, $\sigma_i$, is defined as the variance of velocity: $\sigma_i^2 \equiv \overline{{u'_i}^2} = <[u_i(t)-\overline{u}_i]^2>$, where $< >$ represents the average over all particles tracked during this period. 

Fig. \ref{fig:velocity}A shows the average velocity, $\overline{u}_i$, in the silicon channels (filled circles) and the COC channel (empty circles). In the channel with the smallest height (Table \ref{tab:toroids}, SI I), the low aspect ratio, $h/w = 0.25$, does not support a coherent flow. In the larger channels (SI II - IV and COC), average velocities in  the radial and vertical directions are close to zero, while along the azimuthal direction, a mean angular velocity, $\overline{u}_{\theta}$, increases almost linearly with the channel height, $h$.  Fig. \ref{fig:velocity}B shows a similar linear increase with $h$ of all three components of the fluctuation strength $\sigma_i$. Channels SI I and SI II have similar heights (82 $\mu$m vs 107 $\mu$m) but very different aspect ratios (0.25 vs 0.96). Nevertheless, they exhibit the same fluctuation strength ($\approx$ 1 $\mu$m/s) suggesting that the fluctuation amplitude is determined by the minimum confining dimension. 
For a given channel height, irrespective of chaotic and coherent flow, $\sigma_r$, $\sigma_{\theta}$ and $\sigma_z$ have the same magnitude, indicating small-scale isotropc. This behavior is consistent with previous measurements using the same active fluid, but with only a single axis of confinement \cite{fan2021effects}. In that case, fluctuation strength scaled with the smallest confining geometry, but also remained locally isotropic. 

There are major differences in mean velocities and fluctuation strength magnitudes between results here and our previous study \cite{fan2021effects}. In previous results, the confinement is one-dimensional along $z$-axis, meanwhile the system is unconfined at $x-y$ plane. Such geometry does not persist circulation as torroidal channels here, therefore zero mean velocities. At the equivalent channel height $\sim$$100$ microns, 1D confined rectangular chamber shows about three times stronger fluctuation than 3D confined toroidal channel. In other words, the fluctuation activity is more fundamental principle relating to the degree of confinement (1D or 3D) and confining geometry (the smallest dimension). The circulation coherent flows arises from the fluctuation activity if both the confinement shape and aspect ratio meet the criteria. The circulation mean velocity magnitude rises with the fluctuation magnitude. There are also similarities though the differences we explained. In the previous 1D rectangular geometries and current 3D torroidal channels, both systems are locally isotropic. If we focus on the two largest torroidal channels SI IV and COC, $\overline{u}_{\theta}$, $\sigma_r$, $\sigma_{\theta}$ and $\sigma_z$ all lie at comparable magnitudes (Fig. \ref{fig:velocity}A-B, channel height $h \sim 300$ $\mu$m). It implies that such local isotropic, or to be referred as fluctuation amplitude, is not affected by the circulation speed or even the lifetime of active fluids (Fig. \ref{fig:lifetime}B,C).

\subsection{Three-dimensional profile of active flows}
\begin{figure}
\centering
\includegraphics[width=\columnwidth]{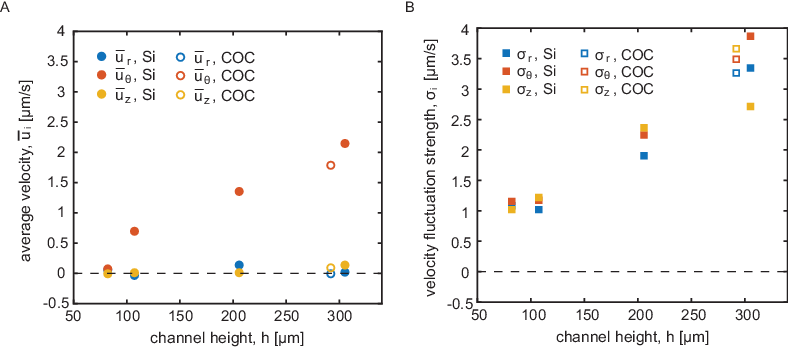}
\caption[Averaged velocity and velocity fluctuation in the active flows.]{(A) Average velocity of the active fluids in multiple channels, averaged during the defined time period, $T = [0 - 2000]$ seconds. Filled circles are data from silicon channels and empty circles are data from the COC channel. (B) Velocity fluctuation strength of the active fluids in multiple channels during the same time period, $T$. Filled squares are data from silicon channels and empty squares are data from the COC channel. }
\label{fig:velocity} 
\end{figure}

\begin{figure}
    \centering
    \includegraphics[width=\columnwidth]{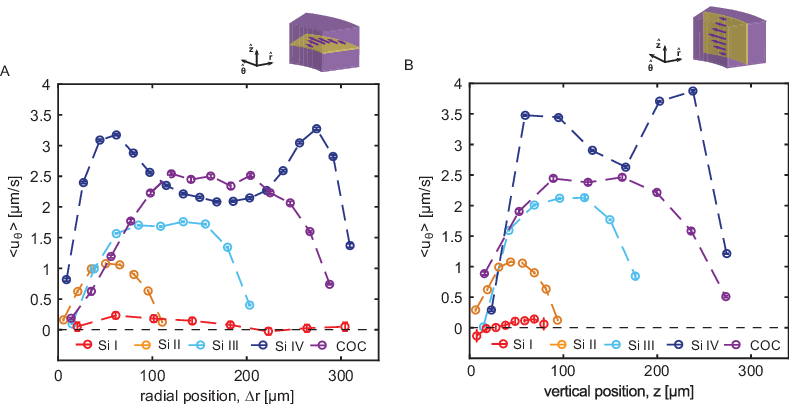}
    \caption{Velocity profiles. (A) Schematic of the "top view" plane and corresponding azimuthal velocity, $u_\theta$, at the mid-height location, as a function of radial position, $r$. (B) Schematic of the "side view" plane and corresponding azimuthal velocity, $u_\theta$, at the mid-radial location, as a function of vertical position, $z$.}
    \label{fig:profiles}
\end{figure}

To visualize the three-dimensional flow structure, we define two averaging strategies: ``top view'' (Fig.~\ref{fig:profiles}A) and ``side view'' (Fig.~\ref{fig:profiles}B). In the top view, the angular velocity vectors are averaged along the azimuthal axis, $\theta$, and over the vertical axis, $z$, to obtain velocity profiles as a function of the radial position, $\Delta r$, where $\Delta r = r - r_{in}$. In the side view, we average over $r$ and $\theta$ to obtain profiles as a function of $z$. Consistent with the global averaged velocities (Fig. \ref{fig:velocity}A), the angular velocity for the chaotic flow system (SI I) hovers around zero (Fig. \ref{fig:profiles}A-B, red symbols). In the channels that support a coherent flow, two regimes of velocity profiles are observed: a ``Poiseuille-like'' profile which reaches the maximum velocity near the channel center,  and a ``peak-separated'' profile where two internal maxima are observed. The 3D velocity profiles in the smaller silicon channels (SI II and III) and in the COC channel are all Poiseuille-like (Fig. \ref{fig:profiles}A-B, orange, cyan and purple symbols). Meanwhile, in the largest silicon channel (SI IV), which has dimensions similar to those of the COC channel, we see four separated peaks, one in each corner of the channel, and the average velocity drops to a local minimum near the channel center (Fig. \ref{fig:profiles}A-B, dark blue symbols). The amplitude of the peak velocity is strongly dependent on the channel geometry. Coherent flows at higher speeds are consistently observed in the larger channels (Fig. \ref{fig:velocity}A and Fig. \ref{fig:profiles}A-B).

In all cases, for the channels with aspect ratio approximately one, the coherent flow maintains this four-fold symmetry regardless of the channel size, material (SI or COC), or surface chemistry. Regardless of the chaotic or coherent dynamical state of the active flow, the averaged azimuthal velocity near the confining surface show very small magnitudes. We did not have enough resolution to visualize the microtubules. Therefore listing two hypotheses. The flows near boundaries could be statistical zero, which supports a ``no-slip'' boundary condition. Or, the near boundary active fluids are similar to the active fluid water-oil surfaces with a thin nematic layer of microtubule bundles \cite{chen2021flow}. In this second hypothesis, the nematic layer introduces small extensile flows near the boundary.

\section{Discussion}
\subsection{Scaling of coherent flows}
\begin{figure}
\centering
\includegraphics[width=\columnwidth]{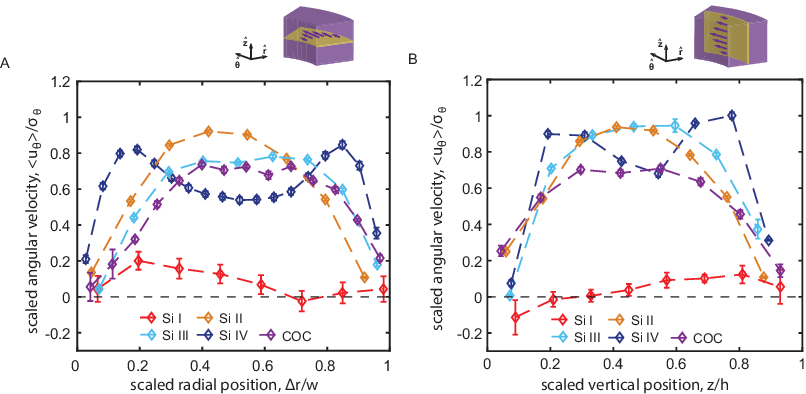}
\caption[Scaled three-dimensional velocity profiles of active flows in different confining geometries.]{\label{fig:scale_profile} 
(A) Scaled velocity profile as a function of radial position. The radial position is normalized by corresponding channel width. (B) Scaled mean angular velocity profile as a function of vertical position. The vertical position is normalized by corresponding channel height.}
\end{figure}

One simple way to proceed is to scale the three-dimensional profiles (Fig. \ref{fig:profiles}) by the amplitude of the velocity fluctuations (Fig. \ref{fig:velocity}B), and to scale the radial and vertical axes by the channel width and height. Doing so, we define a normalized velocity: $\tilde{u}_\theta(\tilde{\Delta r}, \tilde{z}) = {\overline{u}_{\theta}}/{\sigma_{\theta}}$, where $\tilde{\Delta r} = \Delta r / w$ and $\tilde{z} = z / h$. Remarkably, the normalized velocity profiles from the three ``Poiseuille-like'' cases (SI II, SI III, COC) demonstrate an approximately universal structure (Fig. \ref{fig:scale_profile}A-B, orange, cyan, purple) regardless of the channel size. The peak-separated case (SI IV) exhibits a maximum scaled velocity with the same magnitude as the Poiseuille-like cases, $\tilde{u}_\theta = {\cal O}(1)$, although the four peaks are located away from the channel center, at approximately $\tilde{\Delta r} \sim 0.2, 0.8$ and $\tilde{z} \sim 0.2, 0.8$ (Fig. \ref{fig:scale_profile}A-B, dark blue). 

\subsection{Peak-splitting velocity profiles}
Our observations are consistent with previous velocity profiles reported by \cite{wu2017transition} who showed ``Poiseuille-like'' profiles with angular velocities peaking around $1$ $\mu$m/s in a COC torroidal channel with height around $150$ $\mu$m, and width around $100$ $\mu$m, and $2$ to $2.5$ $\mu$m/s in similarly shaped channels with width ranging from $200$ to $300$ $\mu$m. Here, we present similar profiles with maxima around $1$ $\mu$m/s (Fig. \ref{fig:profiles}A, SI II), $1.6$ $\mu$m/s (Fig. \ref{fig:profiles}A, SI III), and $2.5$ $\mu$m/s (Fig. \ref{fig:profiles}A, COC). More recent experiments \cite{chen2021flow} have reported on the velocity profiles and surface interactions in active droplets confined by different thickness of oil layers. Although the surface interaction and confining geometries are completely different from those considered here, the active systems also tend to show ``Poiseuille-like'' velocity profiles. 

However, neither of these related studies \cite{wu2017transition,chen2021flow} were able resolve the volumetric structure of velocity field, and for this reason may not have been able to detect this peak splitting. This exceptional case (Fig. \ref{fig:profiles}A-B SI IV, dark blue) shows separated velocity peaks, indicating that the active system bears four local maxima in the three-dimensional velocity field. This observation was not seen in previous studies using the same or similar kinesin-driven microtubules system \cite{wu2017transition, chen2021flow}, and is only seen in this one experimental configuration, and its presence was confirmed through repeated measurements. The flow in the similarly-sized COC channel exhibits comparable average velocities and fluctuation strength (Fig. \ref{fig:velocity}A-B), but without the peak-splitting.  

One obvious difference between COC and SI IV channels is the channel material and the chemical treatment of the surfaces. However, the smaller silicon channels (SI II,  III) have the same surface treatment, and do not exhibit the split velocity profiles, suggesting that the phenomenon can emerge in channels with the SI-surface treatment but also only in larger systems. This implies that when the system size exceeds some maximum internal large-scale correlation length scale, the velocity profile transitions to a more ``boundary layer'' type of flow - dominated by the ``no-slip'' condition imposed by the nearest wall, but less influenced by the presence of the opposite wall. This explanation is supported by the observation of a maximum correlation length of approximately 400 $\mu$m that was supported in confined kinesin-microtubule systems \cite{fan2021effects}. 

One factor that might contribute to the difference between the profiles in the large SI-IV and COC channels is the observation that the flow in the silicon channels is more long-lived and robust than the flow in the COC channel. In the large silicon channel, the velocity is relatively constant, with an average velocity of 2 $\mu$m/s throughout the 2000 seconds recording window (Fig~\ref{fig:lifetime}B). In contrast, by the end of the 2000 seconds window, the velocity in the COC channel has already started to decline (Fig~\ref{fig:lifetime}C). Although we are not sure how this might affect the velocity profiles, the shorter lifetime in the COC channel reinforces the idea that the surface treatment is a critical feature of the flow. Despite this unsolved mystery, we nevertheless choose to present the full data set from both COC and SI channels, rather than to arbitrarily discard data without good reason.

This peak-separation has also been observed in other active matter systems. Specifically in confined bacterial suspensions, \citet{wioland2016directed} observed that as the confining channel width increases, the velocity profile of the bacterial system also transitions from ``Poiseuille-like'' to separated peaks. In such bacterial suspensions, active vortices with a critical size were captured. One interesting question to raise is whether the local minimum at the center of the channel would reach zero or even negative values between the peaks in the average velocity field, similar to those observed in the bacterial system \cite{wioland2016directed}. 
Addressing this question would require much larger channel dimensions and a re-designed optical measurement system to enable coverage of the greatly enlarged measurement volume. Unfortunately, this lies outside the scope of this current work.

\subsection{Large-scale coherence}
The coherent flows have remarkably large scale, compared to the average length of microtubule ($1$ $\mu$m \cite{wu2017transition}). As mentioned above, with the same active system \citet{fan2021effects} found that a characteristic correlation length scale increases from $50$ $\mu$m to $\sim 400$ $\mu$m as the one-dimensional confining boundary retreats. Here, the active fluid is strongly confined and the velocity profiles reveal coherent structures that span in the confining geometry from $100$ $\mu$m to $300$ $\mu$m with a certain aspect ratio. In other active materials, large-scale coherence has been reported as ``cytoplasmic streaming'' or ``rotational flows''. Cytoplasmic streaming \cite{theurkauf1994premature, serbus2005dynein, schroeder1985spiral} is generated by molecular motors walking on cytoskeletal filaments in the presence of organelle bound in living cells. Due to the restrictions from organelle or cell boundary, directed flow of cytosol shows a characteristic length scale around tens to a hundred micrometers. The self-organized rotating flows in bacterial suspensions, caused by geometrical asymmetry or dimension variation in confinements, show characteristic length scale around tens to a hundred micrometers as well \cite{di2010bacterial, sokolov2010swimming, lushi2014fluid, kim2008microfluidic, you2018geometry}. In other artificially purified cytoskeletal extracts, wall-induced rotational flows emerge in a wide range of confining geometries, from tens to several hundred micrometers \cite{ndlec1997self, schaller2010polar, suzuki2017spatial}. Though our observation and above reports share very similar circulating pattern, a major difference exists. Coherent flows presented here are intrinsically three-dimensional phenomena, whereas, above reports are typically quasi-2D phenomena.

\section{Conclusions}
In summary, we have measured the three-dimensional structure of the coherent flows observed in confined kinesin-driven microtubule systems. In the geometries that support coherent flows (aspect ratios $\sim 1$), the coherent azimuthal flow structures are found to scale with the fluctuation strength, which in turn scales with the confinement length, independent of the surface chemistry. We found that the lifetime, the velocity profiles and the structure of the coherent flows are affected by the confinement materials and/or the coatings on the confining boundaries. The COC confinement with pluronic coatings exhibits a single velocity peak in the center of the channel, and a shorter activity lifetime. In contrast, the silicon confinement with polyacrylamide coatings shows longer lifetime and develops separated peaks at the largest dimension tested. Further experimental investigations are demanded to study the two hypothesis of lifetime variation and observed ``peak-separated'' flow structures - the adsorption on microtubules from silicon surfaces or the density distribution of motor clusters. An interesting proposal would be to repeat the same measurements with fluorescently-labelled kinesin-streptavidin clusters and fluorescent particles, which will reveal the motion of motor clusters and the dynamics of the active flows. However this is beyond the scope of our current project capabilities and will be left to a later time.

There are more interesting open questions. Would the flow near the channel center slow down to ``chaotic'' level in a large-enough silicon channel? This would require a larger field of view in the microscope system with equivalent imaging resolution. Does the observed $50$ $\mu$m - $100$ $\mu$m distance to the coating silicon wall indicate any characteristic length in the active microtubule system? Would that be a different length scale with respect to different surface conditions? To explore these questions, a high resolution 3D microscope system, for example a confocal miscroscope, is highly recommended to be capable of scanning the fluorescent microtubule networks.  

\section{Acknowledgments}
We are most grateful to Tom Powers and Kunta Wu for their insightful comments that have contributed to the analysis of this data, and to Seth Fraden and Zvonimir Dogic for their support and inspiration during the entire course of this research project.  The work was supported by NSF-MRSEC-1420382, NSF-1336638, and NSF-MRSEC-2011486.

% \begin{thebibliography}{9}
% \input PRFv1.bbl
% \end{thebibliography}
\bibliography{structure}

\clearpage
\clearpage
\newpage
{\centering {\Large \textbf{Supplementary Information}}}
\setcounter{page}{1}
\setcounter{figure}{0}
\setcounter{section}{0}
\renewcommand\thefigure{Supp.\arabic{figure}} 
\renewcommand\thesection{Supp.\arabic{section}} 
\renewcommand\thesubsection{Supp.\arabic{section}.\arabic{subsection}}

\section{Materials and Methods}
Microtubules were polymerized from tubulin monomers, which were labeled by Alexa Fluor 568 dye (Thermal Fisher Scientific, A20003, excitation/emission: $578$ nm/$603$ nm) during the purification process. Motor clusters were obtained by mixing kinesin motors with streptavidin tetramers (Invitrogen, S-888). Pluronic micelles F127 (Sigma, P2443) in aqueous active systems force microtubule to bundle by volume exclusion. The ATP regeneration system maintained a steady ATP concentration in assembled system to stabilize motions of molecular motors. For active flow measurement, we dispersed $\sim$0.003$\%$ $(v/v)$ fluorescent particles with Fluo-Max Blue dye ($1$ $\mu$m diameter, Thermal Fisher Scientific, B0100, excitation/emission: $412$ nm/$447$ nm) in the assembled kinesin-driven active fluids.

The active fluids were confined in a series of toroidal channels of width $w$ and height $h$. All the channels had the same inner radius, $r_{in} = 1$ mm (Figure \ref{fig:experiment}). Dimensions of all of the the channels tested are shown in Table~\ref{tab:toroids}. Channels were manufactured either by etching silicon wafers (University Wafer, 444 and 452) or by CNC-machining cyclic olefin copolymer plates (TOPAS Advanced Polymers 5013L-10). Wafers were etched using an inductively-coupled plasma reactive ion etching (RIE) system (SPTS Technologies, Model LPX-ICP), following the Bosch process \cite{tilli2015handbook}.  In order to establish a consistent surface condition, the fabricated silicon channels were coated with polyacrylamide to prevent protein adhesion. This treatment was used on all surfaces that contacted samples, such as microscope cover glass (Fisher scientific, 12-542C, $170$ $\mu$m thickness). In order to assess the influence of different surface conditions, a toroid, with similar dimensions to a silicon channel (Si IV), was CNC-machined (MDA Precision LLC. Model V8-TC8 3-axis) from cyclic olefin copolymer (COC) plates (TOPAS Advanced Polymers, 8007). The COC channel was coated by pluronic F-127 (Sigma, P2443). As with the silicon channels, the pluronic treatment was also used used on a COC film (TOPAS Advanced Polymers, 5013, $101.6$ $\mu$m thickness, refractive index $1.5$), which was used to seal the samples.

The active fluid was loaded into the channel and sealed by bolting together a microfluidic device (Figure \ref{fig:experiment}b) which included, from bottom to top, a sample holder to fit the motorized microscope stage (Prior Scientific, H101A), a treated cover (microscope cover glass or COC film), a treated channel (silicon channel or COC channel), a polydimethylsiloxane (PDMS) layer, $\sim$1 mm thick, to provide stress relief and an aluminum plate. The assembled package could prevent the sample from drying out for more than a week. 

\subsection{Imaging and three-dimensional scanning}
The test cell was mounted on an epi-fluorescent microscope (Nikon TE 200 eclipse). The particles and microtubules were illuminated with a high power LED source (Thorlabs, DC4100) at 
%two-bandwidth optical components, an air immersion objective (Nikon CFI Plan Apo VC $20\times$, WD = 1 mm, NA = 0.75) held by a piezo z-drive (Physik Instrumente, Model P-725), an optical splitter (Cairn Research, Optosplit II) and an sCOMS camera (PCO edge 5.5). The microscope and associated equipment are contained inside a chamber (World Precision Instruments), which maintains the temperature of the experimental environment at $25^{\circ}$C $\pm0.3$. 
$405$ nm and $565$ nm. The excitation light, refined using a two-bandwidth excitation filter (Chroma Technology, 59003x, $395/30$ nm and $575/30$ nm), was directed through an air immersion objective (Nikon CFI Plan Apo VC $20\times$, WD = 1 mm, NA = 0.75) mounted on a piezo objective drive (Physik Instrumente, Model P-725), through the floor of the test cell and into the test chamber. The fluorescently-labelled microtubules and fluorescent tracer particles in the sample emit light at red and blue wavelengths respectively. The emission light was passed through a two-bandwidth dichroic mirror (Chroma Technology, 59003bs, transmission in $460/30$ nm and $630/60$ nm) and was directed to an optical splitter which separates the two colors using a dichroic mirror (Chroma Technology, $565$dcxr). Each light stream was passed through a narrow band filter (Chroma Technology, ET$450/50$m and D$620/60$m) and projected onto two halves of an sCMOS camera (PCO edge 5.5). Although the full sensor size of the camera is $2560$ pixels $\times$ $2160$ pixels, limitations with the windows of the optical splitter result in a final image of $2500$ pixels $\times$ $1300$ pixels. The exposure time of the sensor was $25$ milliseconds. 

With the channel size and imaging system described the field of view was approximately at around $20^{\circ}$ portion of the torroidal channel (Fig.\ref{fig:experiment}b).

\subsection{Volumetric scanning}
The focal plane was scanned over a number of different heights (Table~\ref{tab:toroids}) equally-spaced approximately $30$ $\mu$m to $70$ $\mu$m apart based on the channel height and Z-scanning position numbers. From this $z$-stack a quasi-steady volumetric snapshot of the particles could be assembled. The objective was moved to its location during $50$ ms, and held in place for $50$ ms before each image was acquired (the moving and settling time of the piezo drive is less than $100$ ms). In this manner a complete z-scan was achieved at 1 Hz for scanning position $M=3$ or $M=5$ and 2 Hz for scanning position $M=10$. Given that the average particle velocity is approximately 2 $\mu$m/s, the slew error between different slices is minimal and can be ignored. 

\subsection{Particle localization and tracking}
The three-dimensional positions are acquired from each image by detecting the Airy disks associated with out-of-focus particles \cite{speidel2003three, wu2005three, park2006three, taute2015high}. The radius of the Airy disk reflects the $z$-position of an out-of-focused particle and the center of the Airy disk determines the position in the horizontal plane. Therefore, three-dimensional particle tracking is eventually transformed to a circle detection problem. We use a well-known method called the Circle Hough Transform (CHT) \cite{ballard1987generalizing}. The detection code, written in MATLAB, follows established pattern recognition concepts \cite{nasrabadi2007pattern}, the CHT algorithm \cite{afik2015robust} and parallel processing method (MATLAB Imaging Processing Toolbox). A reference library, which calibrates the Airy disk radius with the $z$-location of a particle, was obtained by scanning a $1$ $\mu$m diameter fluorescent particle, immobilized in $1\%$ $(w/w)$ agarose gel, over a range of $100$ $\mu$m along $z$-axis at a step of $0.1$ $\mu$m. Scanned images are stacked to resolve a light cone (Figure \ref{fig:experiment}c) of the fluorescent particle \cite{taute2015high}. After the CHT, centers and radii of diffracted particle in each frame are acquired. Instant particle positions are calculated from centers and radii with respect to the reference library and the focal plane position. 

From the time sequence of particle positions, particles were tracked using the ``nearest neighborhood'' method \cite{adrian1991particle} and local fluid velocities computed using finite differences. Average, $\bar{u}_i$ and RMS, $\sigma_i$, quantities were computed by grouping particles at different $r$ and $z$ locations into spatial bins and averaging the velocities of all particles in each bin over 2000 seconds (Fig~\ref{fig:velocity}). The bin sizes were $12$ to $20$ $\mu$m in the $r$-direction and $12$ to $38$ $\mu$m in the $z$-direction. Since the azimuthal direction, $\theta$, is homogeneous, the velocities were averaged in the azimuthal direction.

\section{Chaotic and Coherent flow regimes}
\begin{figure}
    \centering
    \includegraphics[width=\columnwidth]{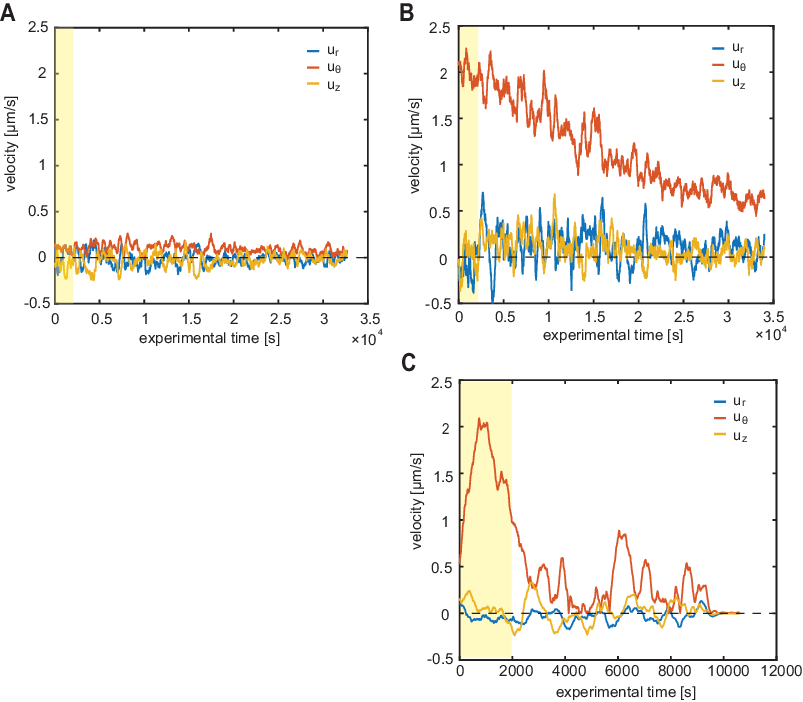}
    \caption{(A) Velocity of chaotic flows in the silicon channel (SI I) as a function of experimental time. (B) Velocity of coherent flows in the silicon channel (SI IV) as a function of experimental time. (C) Velocity of coherent flows in the COC channel. All velocity values are locally smoothed over a neighbor period of $200$ seconds. The highlighted area defines a 2000 s time period, $T = [t_s, t_e]$, over which temporal averaging takes place.}
    \label{fig:lifetime}
\end{figure}

Figure \ref{fig:lifetime} illustrates three typical experiments of active flows among $10$ data sets in total recorded from $5$ different channels. Radial and vertical velocity components always fluctuate around zero. The existence of coherent flows is determined by the angular velocity. Figure \ref{fig:lifetime}A shows chaotic flows in the silicon channel (Si I) with a small aspect ratio, $h/w = 0.25$. The angular velocity fluctuates around zero as a function of experiment time, implying that no long-ranged coherence is observed. Figure \ref{fig:lifetime}B shows coherent flows in the silicon channel (Si IV) with a large aspect ratio, $h/w=0.95$. The angular component of velocity initiates around $2$ $\mu$m/s and decays monotonically towards $0.5$ $\mu$m/s with chemical energy consumption, indicating the existence of long-term coherence for over $9$ hours. Angular velocities, measured from other silicon channels (Si II and Si III), present similar decays as a function of time with smaller initial speeds. 
%Figure \ref{fig:lifetime}C shows coherent flows in the COC channel. The lifetime of active flows shortens to $3$ hours and angular velocity component illustrates an exponential decay with damping as a function of experiment time. 
The yellow-highlighted areas both plots define a time period, $T=2000$ seconds, for the temporal averaging of velocity.

\end{document}